\documentclass[twocolumn,tighten]{aastex631}
\usepackage{graphicx}
\usepackage{amsmath}
\usepackage{natbib}
\usepackage{amssymb}
\usepackage{ulem}
\usepackage{sidecap}
\usepackage{hyperref}

\def\Msun{\ensuremath{\mathrm{M}_\odot}}

\shorttitle{High-z GRBs \& JWST}
\shortauthors{Matsumoto et al.}

\begin{document}

\title{Probing the Origin of the Star Formation Excess Discovered by JWST through Gamma-Ray Bursts}

\author[0000-0002-9350-6793]{Tatsuya Matsumoto}
\affil{Department of Astronomy, Kyoto University, Kitashirakawa-Oiwake-cho, Sakyo-ku, Kyoto, 606-8502, Japan}
\affil{Hakubi Center, Kyoto University, Yoshida-honmachi, Sakyo-ku, Kyoto, 606-8501, Japan}

\author[0000-0002-6047-430X]
{Yuichi Harikane}
\affil{Institute for Cosmic Ray Research, The University of Tokyo, 5-1-5 Kashiwanoha, Kashiwa, Chiba 277-8582, Japan}

\author[0000-0003-2611-7269]{Keiichi Maeda}
\affil{Department of Astronomy, Kyoto University, Kitashirakawa-Oiwake-cho, Sakyo-ku, Kyoto, 606-8502, Japan}

\author[0000-0002-3517-1956]
{Kunihito Ioka}
\affil{Center for Gravitational Physics and Quantum Information, Yukawa Institute for Theoretical Physics, Kyoto University, Kyoto 606-8502, Japan}

\begin{abstract}
The recent observations by the James Webb Space Telescope (JWST) have revealed a larger number of bright galaxies at $z\gtrsim10$ than was expected. The origin of this excess is still under debate, although several possibilities have been presented. We propose that gamma-ray bursts (GRBs) are a powerful probe to explore the origin of the excess and, hence, the star and galaxy formation histories in the early universe. Focusing on the recently launched mission, Einstein Probe (EP), we find that EP can detect several GRBs annually at $z\gtrsim10$, assuming the GRB formation rate calibrated by events at $z\lesssim6$ can be extrapolated. Interestingly, depending on the excess scenarios, the GRB event rate may also show an excess at $z\simeq10$, and its detection will help to discriminate between the scenarios that are otherwise difficult to distinguish. Additionally, we discuss that the puzzling, red-color, compact galaxies discovered by JWST, the so-called ``little red dots'', could host dark GRBs if they are dust-obscured star forming galaxies. We are eager for unbiased follow-up of GRBs and encourage future missions such as HiZ-GUNDAM to explore the early universe. 
\end{abstract}


\section{Introduction}
\label{sec:introduction}
Gamma-ray bursts (GRBs) are among the most luminous explosions in the universe, allowing them to be detected from large distances \citep{Lamb&Reichart2000,Ciardi&Loeb2000,Gou+2004}. This characteristic makes GRBs invaluable as probe of the distant universe. In particular, long GRBs, which have a nominal duration of longer than $2\,\rm s$ \citep{Kouveliotou+1993}, are produced by the collapse of massive stars \citep[e.g.,][]{Woosley1993,MacFadyen&Woosley1999} and thus serve as powerful tools for investigating star formation processes throughout cosmic history \citep{Totani1997,Wijers+1998,Krumholz+1998,Mao&Mo1998,Blain&Natarajan2000,Porciani&Madau2001}. Actually, the detection of high-$z$ GRBs at $z\gtrsim6$ by \textit{Swift} \citep[see,][for a review]{Salvaterra2015} has prompted active researches into the potential of GRBs to explore the history of cosmic star formation \citep{Price+2006,Daigne+2006,Le&Dermer2007,Guetta&Piran2007,Chary+2007b,Kistler+2008,Li2008,Yuksel+2008,Kistler+2009,Wang&Dai2009,Qin+2010,Wanderman&Piran2010,Ishida+2011,Robertson&Ellis2012,Salvaterra+2012,Wang2013}, reionization (\citealt{Kawai+2006,Totani+2006,Gallerani+2008,McQuinn+2008,Greiner+2009,Patel+2010,Chornock+2013,Totani+2014,Hartoog+2015,Fausey+2024}, see also \citealt{MiraldaEscude1998}), and the first generation of stars \citep{Bromm&Loeb2006,Meszaros&Rees2010,deSouza+2011,Suwa&Ioka2011,Toma+2011,Nagakura+2012,Nakauchi+2012,Kashiyama+2013b,Matsumoto+2015,Matsumoto+2016,Kinugawa+2019}.

Recently, significant progress has been made in observations of the high-$z$ universe with the advent of the James Webb Space Telescope (JWST). JWST not only broke the record for the most distant galaxy observed but also revealed that there are more abundant bright galaxies than previously expected \citep[e.g.,][]{Finkelstein+2022c,Finkelstein+2023,Finkelstein+2024,Naidu+2022,Adams+2023,Adams+2024,Bouwens+2023,Bouwens+2023b,Castellano+2023,Donnan+2023,Donnan+2023b,Donnan+2024,Harikane+2023,Harikane+2024,Harikane+2024b,PerezGonzalez+2023,McLeod+2024,Robertson+2024}. Various ideas have been proposed to explain this ``JWST excess'' in the UV luminosity function, such as active star formation \citep{Dekel+2023,Li+2023}, top-heavy initial mass function \citep{Inayoshi+2022,Steinhardt+2023}, and even a flaw in the cosmological model \citep{Parashari&Laha2023}; however, the cause remains unclear, and it has become one of the topics of active debate \citep[see discussions in][]{Harikane+2023,Harikane+2024,Harikane+2024b}.

In this letter, we explore the detectability of high-$z$ GRBs and their potential to elucidate the origin of the excess discovered by JWST. In particular, we focus on the observational prospect of the \textit{Einstein Probe} \citep[EP,][]{YuanWeimin+2022}\footnote{\texttt{https://ep.bao.ac.cn/ep/}}, which was launched in January 2024 and has begun its observations. EP has a sensitive soft X-ray detector Wide X-ray Telescope (WXT), being advantageous for observing high-$z$ GRBs. In fact, a GRB at $z=4.859$ was already detected \citep{Gillanders+2024,Liu+2024,Levan+2024,Ricci+2024}. We find that depending on potential origins of the JWST excess, the GRB formation rate can have different behaviors around $z\gtrsim10$, and its detection by EP or future GRB missions will clarify the cause of the JWST excess.

The paper is organized as follows. In Sec.~\ref{sec:method} we describe the method to calculate the event rate of GRBs. In Sec.~\ref{sec:rate}, various possibilities on the GRB event rates at $z\gtrsim10$ are discussed along with the origin of the JWST excess. We present our results in Sec.~\ref{sec:result}, and summarize our findings in Sec.~\ref{sec:summary}. Throughout this Letter, we assume $\Lambda$CDM cosmology and use the cosmological parameters of $H_0=67.4\,\rm km\,s^{-1}\,Mpc^{-1}$, $\Omega_{\rm m}=0.315$, and $\Omega_{\Lambda}=0.685$ \citep{PlanckCollaboration2020_parameter}.

\section{Event rate of high-$z$ GRBs}\label{sec:method}
The number of GRBs detected by a detector during an observation time $\Delta t_{\rm obs}$ and for a redshift range of ($z,\,z+dz$) is calculated by \citep[e.g.,][]{Bromm&Loeb2006,deSouza+2011}
\begin{align}
\frac{dN_{\rm GRB}}{dz}&=\Psi_{\rm GRB}^{\rm obs}\frac{\Delta t_{\rm obs}}{1+z}\frac{dV}{dz}\ ,
    \label{eq:dNdz}
\end{align}
where $\Psi_{\rm GRB}^{\rm obs}$ is the observed comoving event rate of GRBs, that is the number of observable GRBs per comiving volume and time, discussed in the next paragraph, and the cosmological volume element is given by
\begin{align}
\frac{dV}{dz}&=\frac{4\pi c d_{\rm L}^2}{1+z}\left|\frac{dt}{dz}\right|=\frac{4\pi c d_{\rm L}^2}{(1+z)^2H_0\sqrt{\Omega_{\rm m}(1+z)^3+\Omega_\Lambda}}\ .
    \label{eq:dVdz}
\end{align}
Here $c$ and $d_{\rm L}$ are the speed of light and the luminosity distance, respectively.

The comoving event rate of GRBs observed by the detector covering a solid angle $\Omega$ in the sky is given by
\begin{align}
\Psi_{\rm GRB}^{\rm obs}(z)=\frac{\Omega}{4\pi}\eta_{\rm beam}\Psi_{\rm GRB}(z)\int_{L_{\rm min}(z)}^{\infty}\frac{dn}{dL}dL\ ,
    \label{eq:Psi^obs_GRB}
\end{align}
where $\eta_{\rm beam}$ is the beaming factor giving a fraction of on-axis events, $\Psi_{\rm GRB}(z)$ is the intrinsic comoving GRB formation rate, $L_{\rm min}(z)$ is the minimal GRB luminosity to trigger the detector, and ${dn}/{dL}$ is the normalized luminosity function (LF, $\int \frac{dn}{dL}dL=1$). In this Letter we call an isotropic equivalent gamma-ray luminosity as a luminosity for simplicity. For a jet with a half-opening angle $\theta_{\rm j}$, the beaming factor is given by
\begin{align}
\eta_{\rm beam}=1-\cos\theta_{\rm j}\simeq\frac{\theta_{\rm j}^2}{2}\simeq0.005\,\left(\frac{\theta_{\rm j}}{0.1}\right)^2\ ,
    \label{eq:eta_beam}
\end{align}
where $\theta_{\rm j}=0.1$ is an observationally motivated value \citep[e.g.,][]{Frail+2001,Goldstein+2016}. Our results will linearly depend on the beaming factor.

The minimal luminosity in Eq.~\eqref{eq:Psi^obs_GRB} is calculated by equating a flux of a prompt GRB emission with the detector's sensitivity. To obtain the former one, we assume that the prompt emission has the Band spectrum \citep{Band+1993}:
\begin{align}
N(E)=A\begin{cases}
\left(\frac{E}{100}\right)^{\beta}\left(\frac{(\alpha-\beta)E_0}{100e}\right)^{\alpha-\beta}&: E<(\alpha-\beta)E_0\ ,\\
\left(\frac{E}{100}\right)^{\alpha}\exp\left(-\frac{E}{E_0}\right)&: (\alpha-\beta)E_0<E\ ,\\
\end{cases} 
    \label{eq:Band_func}
\end{align}
where $A$ and $e$ are a normalization and the Napier's constant, respectively. The photon energy is measured in a unit of keV. We fix the low and high energy power-law indices to typical values of $\alpha=-1$ and $\beta=-2.3$ \citep{Preece+2000,Kaneko+2006}, respectively. $E_{\rm 0}$ is the cut-off energy, which is related to the peak energy in the energy spectrum, $E^2N(E)$, as $E_{\rm p}=(\alpha+2)E_0$. The normalization $A$ is obtained for a given luminosity, which is defined by integrating the specific luminosity over $1\,\rm keV$ to $10\,\rm MeV$ in the rest frame:
\begin{align}
L\equiv\int_{1{\,\rm keV}}^{10{\,\rm MeV}}L_{E^\prime}dE^\prime=4\pi d_{\rm L}^2\int_{\frac{1{\,\rm keV}}{1+z}}^{\frac{10{\,\rm MeV}}{1+z}}EN(E)dE\ ,
\end{align}
where $E^\prime$ is the photon energy at the rest frame. We use the relations between the energy flux and specific luminosity
\begin{align}
EN(E)=\frac{(1+z)L_{E^\prime}}{4\pi d_{\rm L}^2}\ ,
\end{align}
and between the photon energies at the observer and rest frames $E=E^\prime/(1+z)$ in the second equality. The peak energy is estimated by assuming its empirical correlation with luminosity, the so-called Yonetoku relation \citep{Yonetoku+2004}
\begin{align}
\frac{L}{10^{52}{\,\rm erg\,s^{-1}}}\simeq2\times10^{-5}\left[E_{\rm p}(1+z)\right]^2\ ,
    \label{eq:Yonetoku}
\end{align}
where again the photon energy is measured in a unit of keV. Once the spectral parameters are specified for given $z$ and $L$, the flux of the GRB observed by a detector is calculated by integrating the flux, $EN(E)$, over the detector's energy range. The minimal luminosity is defined by the luminosity giving the same flux as the detector's sensitivity. In Table~\ref{table:detector}, we show the limiting sensitivity and other properties of \textit{Swift} BAT and EP WXT as well as new and future missions. Conservatively we set the fiducial value of EP sensitivity as $\simeq10^{-10}\,\rm erg\,s^{-1}\,cm^2$ (obtained by an exposure time of $\simeq100\,\rm s$), while its maximal sensitivity is $\simeq2.6\times10^{-11}\,\rm erg\,s^{-1}\,cm^2$ achieved for $\simeq1000\,\rm s$ of exposure.

\begin{figure}
\begin{center}
\includegraphics[width=85mm, angle=0, bb =0 0 288 226]{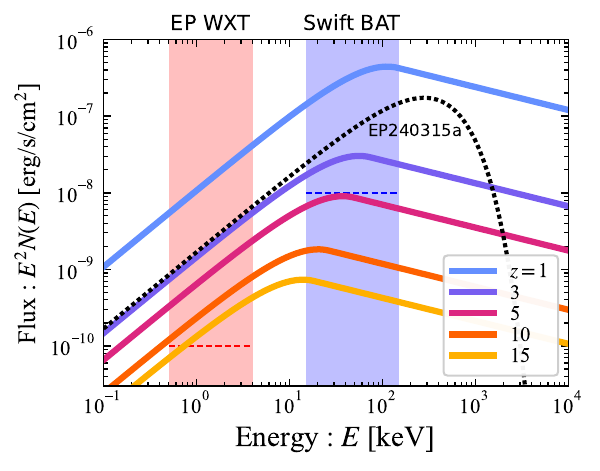}
\caption{Gamma-ray spectra of a GRB with luminosity $L=10^{52}\,\rm erg\,s^{-1}$ at different redshifts. The red and blue shaded regions and dashed lines represent the energy bands and sensitivities of EP WXT and \textit{Swift} BAT, respectively. The black dotted curve shows the spectrum of EP240315a with  $L\simeq1.2\times10^{53}\,\rm erg\,s^{-1}$ and $z\simeq4.9$ \citep{Liu+2024}. 
}
\label{fig:flux}
\end{center}
\end{figure}

Figure~\ref{fig:flux} depicts the flux of a GRB with $L=10^{52}\,\rm erg\,s^{-1}$ for different $z$. The shaded regions and horizontal dashed lines represent the energy band and sensitivity of \textit{Swift} BAT and EP WXT, respectively. Flux in the low energy band is not significantly reduced compared to the high energy band for increasing $z$ \citep[see also][]{Ghirlanda+2015,Palmerio&Daigne2021}. This is because the spectrum has a rising shape at low energy, and the spectral peak decreases not only in flux but also in energy for higher redshift (see Eq.~\ref{eq:Yonetoku}). Both effects result in a moderate flux reduction in the low energy band. A similar effect was pointed out by \cite{Ciardi&Loeb2000} for afterglow emissions and observationally confirmed by \cite{Frail+2006} for radio afterglows.

We remark that the power-law spectrum of the Band function in soft energy can be extrapolated to EP's band range. This is supported by the simultaneous detection of EP240315a, a GRB at $z\simeq4.9$ with $L\simeq1.2\times10^{53}\,\rm erg\,s^{-1}$ (black dotted curve in Fig.~\ref{fig:flux}), with \textit{Swift} BAT \citep[see Fig.~2b of][]{Liu+2024}. The recently reported detection of EP240219a with \textit{Fermi} GBM also supports such an extrapolation \citep{Yin+2024}.

\begin{table*}
\begin{center}
\caption{The properites of current and future GRB missions.}
\label{table:detector}
\begin{tabular}{lcccc}
\hline
Detector&Energy range&Field of view ($\Omega$)&Sensitivity&Ref.\\
&[keV]&[str]&[erg/s/cm$^{2}$]\\
\hline
Swift BAT&$15-150$&1.4&$10^{-8}$&\cite{Barthelmy+2005}\\
Einstein Probe WXT&$0.5-4$&1.1&$\sim10^{-10}$&\cite{YuanWeimin+2022}\\
SVOM ECLAIRs&$4-150$&2&$10^{-9}$&\cite{Wei+2016}\\
HiZ-GUNDAM&$0.5-4$&0.5&$10^{-10}$&\cite{Yonetoku+2024}\\
THESEUS SXI&$0.3-5$&0.5&$10^{-10}$&\cite{Amati+2018,Amati+2021}\\
THESEUS XGIS&$2-10000$&2&$10^{-8}$&\cite{Amati+2018,Amati+2021}\\
Gamov Explore LEXT&$0.5-5$&0.75&$10^{-10}$&\cite{White2020}\\
\hline
\end{tabular}
\end{center}
\end{table*}

The intrinsic GRB formation rate, $\Psi_{\rm GRB}(z)$, and LF, $dn/dL$, in Eq.~\eqref{eq:Psi^obs_GRB} are not well understood. In fact, one of the goals of the GRB population study is to determine them by fitting the distribution of observables such as flux, peak energy, and redshift \citep{Wijers+1998,Krumholz+1998,Mao&Mo1998,Blain&Natarajan2000,Porciani&Madau2001,LloydRonning+2002,Firmani+2004,Natarajan+2005,Guetta+2005,Jakobsson+2006,Daigne+2006,Salvaterra&Chincarini2007,Le&Dermer2007,Chary+2007b,Kistler+2008,Li2008,Yuksel+2008,Kistler+2009,Salvaterra+2009,Wang&Dai2009,Qin+2010,Wanderman&Piran2010,Ishida+2011,Virgili+2011,Robertson&Ellis2012,Salvaterra+2012,Wang2013,Sun+2015,Petrosian+2015,Perley+2016b,Lan+2019,Lan+2021,Palmerio&Daigne2021,Ghirlanda&Salvaterra2022}. These population studies commonly parameterize the GRB rate and LF assuming their functional forms and determine them by reproducing the observables. While these functional forms differ for each work, a general conclusion is that if the redshift evolution of the LF is taken into account, the rate and LF are degenerate \citep[e.g.,][]{Salvaterra&Chincarini2007,Qin+2010,Virgili+2011,Salvaterra+2012,Palmerio&Daigne2021,Ghirlanda&Salvaterra2022}. With the current sample, it is still difficult to break the degeneracy. Therefore, in this work we adopt several parameterized GRB rate and LF found by recent representative population studies \citep{Lan+2021,Palmerio&Daigne2021,Ghirlanda&Salvaterra2022} to discuss the detectability of high-$z$ GRBs by EP.

\begin{figure}
\begin{center}
\includegraphics[width=85mm, angle=0, bb=0 0 315 209]{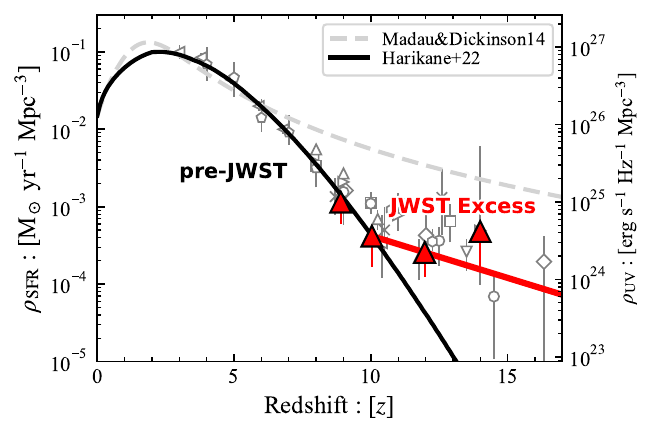}
\caption{The evolution of the cosmic SFR density. The black curve shows the SFR density obtained in the ``pre-JWST'' era. The red triangles denote the lower limit of the SFR density estimated by the spectroscopically-confirmed galaxy sample obtained by JWST in \cite{Harikane+2024b}. The gray open points also show the SFR density obtained by other observations, taken from \cite{Harikane+2024b}. The SFR density shows an excess beyond $z\gtrsim10$, which we call the ``JWST excess" and may be represented by the red solid line. The gray dashed curve shows the SFR density of \cite{Madau&Dickinson2014}, which is frequently used in GRB population studies. The SFR density is estimated by multiplying the conversion factor to the UV luminosity density (on the right axis, see text).}
\label{fig:SFR}
\end{center}
\end{figure}

Before describing the details of the works, we remark on the general nature of the GRB rate. GRB rate can be related to the star formation rate (SFR) since long GRBs are produced by the death of short-lived  massive stars \cite[e.g.,][]{Wijers+1998,Blain&Natarajan2000,Porciani&Madau2001,Yuksel+2008,Kistler+2009,Robertson&Ellis2012}. Here we assume that the GRB formation rate is related with the cosmic SFR density at each redshift by 
\begin{align}
\Psi_{\rm GRB}(z)=\eta_{\rm GRB}(z)\rho_{\rm SFR}(z)\ ,
    \label{eq:Psi_GRB}
\end{align}
where $\eta_{\rm GRB}$ represents the efficiency of GRB formation per stellar mass. Since the GRB formation rate and SFR density have different units, the efficiency has a unit of [$\Msun^{-1}$]. The observations before the advent of JWST (we call ``pre-JWST'') found that the SFR density declines with higher redshift \cite[e.g.,][]{Bouwens+2015,Finkelstein+2015,Harikane+2018,Harikane+2022}. Specifically, as a baseline, we take $\rho_{\rm SFR}$ in \cite{Harikane+2022,Harikane+2024b} as shown in Fig.~\ref{fig:SFR}, whose behavior is understood by structure formation in $\Lambda$CDM cosmology with a constant star formation efficiency. We emphasize that most GRB population studies have adopted the SFR density proposed by \cite{Madau&Dickinson2014} (gray dashed curve) or \cite{Hopkins&Beacom2006}. However, it overshoots the observed SFR density beyond $z\gtrsim6$.

Remarkably, as shown in Fig.\ref{fig:SFR}, the JWST revealed that the SFR density beyond $z\gtrsim10$ is higher than the extrapolation of the pre-JWST results \citep{Finkelstein+2022c,Finkelstein+2023,Finkelstein+2024,Naidu+2022,Adams+2023,Adams+2024,Bouwens+2023,Bouwens+2023b,Castellano+2023,Donnan+2023,Donnan+2023b,Donnan+2024,Harikane+2023,Harikane+2024,Harikane+2024b,PerezGonzalez+2023,McLeod+2024,Robertson+2024}. Note that these JWST studies measure the UV luminosity density (the right axis of Fig.~\ref{fig:SFR}) and convert it to the SFR density by using galaxy SED models. A commonly used conversion factor is calculated for a stellar population with solar metallicity and the Salpeter initial mass function \citep[IMF,][]{Salpeter1955} for $0.1-100\,\Msun$. With a typical conversion factor of $1.15\times10^{-28}\,\rm \Msun\,yr^{-1}/(erg\,s^{-1}\,Hz^{-1})$ \citep[e.g.,][]{Madau&Dickinson2014}, the overabundance of luminous galaxies is translated to an excess of SFR density over the pre-JWST extrapolation. This JWST excess is shown as a red solid line in Fig.~\ref{fig:SFR}, which is obtained by just connecting two data points at $z=10$ and $12$ and extrapolating the line. However, this line could represent a minimal density because the data points are the lower limits. If the JWST excess really reflects an excess of the SFR density, it may also increase the GRB event rate, which is explored in the next section.

In this Letter, we first adopt the GRB formation rate and LF obtained by \cite{Ghirlanda&Salvaterra2022}, hereafter GS22, as a representative (we also do the same calculation for the functions in \citealt{Lan+2021,Palmerio&Daigne2021} and obtained similar results; see Appendix \ref{Appendix:other model}). GS22 analyzed the GRB sample obtained by \textit{Swift} BAT up to 2014 \citep{Salvaterra+2012,Pescalli+2016} to reproduce its statistical properties. They adopted a broken power-law LF defined for $L>10^{47}\,\rm erg\,s^{-1}$ with a redshift-evolving break luminosity:
\begin{align}
\frac{dn}{dL}&\propto\begin{cases}
L^{-p_1}&:\,L\leq L_{\rm b}\ ,\\
L^{-p_2}&:\,L_{\rm b}<L\ ,\\
\end{cases}
    \label{eq:LF_GS22}\\
L_{\rm b}&=L_*(1+z)^{k}\ ,
    \label{eq:LF_GS22_Lb}
\end{align}
and obtained $p_1=0.97$, $p_2=2.21$, $L_*=10^{52.02}\,\rm erg\,s^{-1}$, and $k=0.64$. For the GRB formation rate, they assumed that it has the same functional form as the SFR density of \cite{Madau&Dickinson2014}:
\begin{align}
\Psi_{\rm GRB}=\Psi_{\rm GRB,0}\frac{(1+z)^{q_1}}{1+\left(\frac{1+z}{q_2}\right)^{q_3}}\ ,
    \label{eq:Psi_GRB_GS22}
\end{align}
and obtained $\Psi_{\rm GRB,0}=79\,\rm Gpc^{-3}\,yr^{-1}$, $q_1=3.33$, $q_2=3.42$, $q_3=6.21$. It should be noted that they varied the jet opening angle event by event according to an empirical correlation between the angle and radiated gamma-ray energy. On the other hand, we assume a constant jet opening angle for all events, which may not affect the result significantly given the weak dependence\footnote{This scaling is obtained by combining the so-called Ghirlanda relation $E_{\rm p}\propto (\eta_{\rm beam}E_{\gamma,\rm iso})^{0.7}$ \citep{Ghirlanda+2007}, where $E_{\gamma,\rm iso}$ is the isotropic equivalent gamma-ray energy, Amati relation $E_{\rm p}\propto E_{\gamma,\rm iso}^{0.6}$ \citep{Amati+2002}, and Yonetoku relation of Eq.~\eqref{eq:Yonetoku}.} of $\theta_{\rm j}$ on $L$ suggested by the correlation $\theta_{\rm j}\propto L^{-0.06}$ \citep[but see also][]{LloydRonning+2020}.

\begin{figure}
\begin{center}
\includegraphics[width=85mm, angle=0, bb=0 0 278 286]{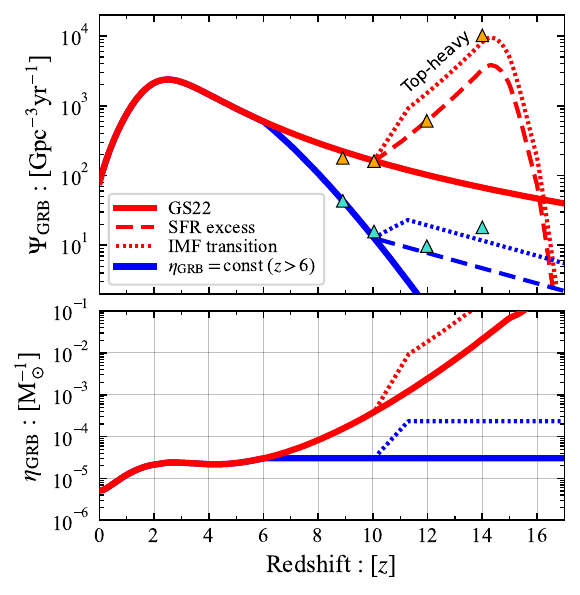}
\caption{({\bf Top}) Intrinsic (beaming-corrected) GRB formation rate for different scenarios. Red and blue solid curves show the rate of \citealt{Ghirlanda&Salvaterra2022} (GS22) and the rate extended beyond $z>6$ in proportion to the pre-JWST SFR. The latter overlaps with the former below $z<6$, and has a constant GRB formation efficiency $\eta_{\rm GRB}$. Dashed curves show the same as the solid ones but boosted in proportion to the SFR excess for $z>10$. Orange and light-blue triangles represent the GRB-rate excess corresponding to the JWST excess (see Fig.~\ref{fig:SFR}) in this scenario. Dotted curves show the case where the JWST excess is caused by a transition of IMF from Salpeter to top-heavy shape over $10\lesssim z \lesssim 11.3$. For the models of the SFR excess (red dashed) and IMF transition (red dotted), $\Psi_{\rm GRB}$ is artificially suppressed for $z\gtrsim14$. ({\bf Bottom}) GRB formation efficiency corresponding to models in the top panel, defined by Eq.~\eqref{eq:Psi_GRB} for the pre-JWST SFR, $\rho_{\rm SFR}$ (black curve in Fig.~\ref{fig:SFR}).}
\label{fig:Psi_GRB}
\end{center}
\end{figure}

Figure~\ref{fig:Psi_GRB} shows the GRB formation rate obtained by GS22 and the corresponding formation efficiency ($\eta_{\rm GRB}$, Eq.~\ref{eq:Psi_GRB}) calculated for ``pre-JWST'' SFR density (Fig.~\ref{fig:SFR}). The efficiency weakly increases for $z\lesssim3$, which is interpreted by the facts that GRBs preferentially occur in low-metal galaxies, and that cosmic metallicity decreases for higher redshift \citep[e.g.,][]{Langer&Norman2006}. We stress that the GRB sample analyzed by GS22 contains bursts only up to $z\lesssim6$, and $\Psi_{\rm GRB}$ beyond $z\gtrsim6$ is not calibrated with observations but just an extrapolation of Eq.~\eqref{eq:Psi_GRB_GS22}. Actually the efficiency increases again for $z\gtrsim6$ to as high as $\eta_{\rm GRB}\sim10^{-2}$, which corresponds to an extreme situation where most massive stars collapse to BH and produce GRB. This may be artificial and only reflect the discrepancy between the SFR formula of \cite{Madau&Dickinson2014} and the observed pre-JWST SFR density. However, we still adopt this formation rate to estimate the high-$z$ GRB event rate as the \textit{most optimistic} model.\footnote{We note that the recent measurement of a low oxygen-to-iron ratio at a galaxy at $z\simeq10.6$ may suggest that collapsars or hypernovae are frequent \citep{Nakane+2024}.} 

A more conservative GRB rate may be obtained by extending $\Psi_{\rm GRB}$ of GS22 from $z\simeq6$ in proportion to the pre-JWST SFR density with a constant $\eta_{\rm GRB}$ as in Eq.~\eqref{eq:Psi_GRB}. This scenario is depicted in a thick blue solid curve in Fig.~\ref{fig:Psi_GRB}.
We remark that the GRB detection rate calculated by this formation rate gives a negligibly small detection rate at $z\simeq8-9$ for \textit{Swift} BAT, while it detected several such GRBs; GRB 090423 at $z=8.2$ \citep{Salvaterra+2009b,Tanvir+2009} and GRB 090429 at $z\simeq9.4$ \citep{Cucchiara+2011}. Therefore, we may regard this scenario as the \textit{most conservative} model.

\section{GRB formation rates and the JWST excess}\label{sec:rate}
Including the above two cases, there are several possibilities on the evolution of $\Psi_{\rm GRB}$ for $z\gtrsim10$. They are motivated by different scenarios of the origin of the JWST excess in the SFR density (Fig.~\ref{fig:SFR}). These scenarios can be summarized as follows: 
\begin{itemize}
\item {\bf Case A: SFR excess.} A real elevation of the SFR can cause the JWST excess. Such an efficient star formation at $z\gtrsim10$ is expected via e.g., the feedback-free \citep{Dekel+2023,Li+2023} and compact star formation \citep[e.g.,][]{Fukushima&Yajima2021}. Note that here the conversion factor from the UV luminosity to SFR does not evolve in time (in contrast to case B discussed below), and the star formation efficiency should be larger than the pre-JWST one found in \cite{Harikane+2022}.\footnote{Such intense star formation may cause radiation-drive winds, which clear up dust from galaxies and play a role to shape the bright end in the UV luminosity function \citep{Ferrara+2023,Tsuna+2023c,Ferrara2024}.} In this case A, the GRB formation rate is also elevated in the same amount as the SFR according to Eq.~\eqref{eq:Psi_GRB} because an IMF does not change. The corresponding GRB formation rates are shown by dashed curves in the upper panel of Fig.~\ref{fig:Psi_GRB} as deviations from GS22 (red solid) and $\eta_{\rm GRB}=\rm{const}$ models (blue solid curve).

\item {\bf Case B: IMF transition.}
A transition of the Salpeter IMF to top-heavy one increases the conversion factor from the UV luminosity to SFR, which may result in the JWST excess without a genuine excess of the SFR \citep{Chon+2022,Inayoshi+2022,Steinhardt+2023}. For example, \cite{Harikane+2023} demonstrated that the conversion factor becomes $\simeq3$ times higher for a top-heavy IMF than Salpeter one (see their Figure~20, and Table 1 of \citealt{Inayoshi+2022} for details).

We construct $\Psi_{\rm GRB}$ in this scenario by assuming that the IMF gradually shifts to a top-heavy shape while the true SFR density traces the pre-JWST one. However, with the same parameters as \cite{Inayoshi+2022}, we find that for the IMF transition alone cannot sustain the JWST excess beyond $z\gtrsim11.3$, and hence an excess of the true SFR is still required (but $\simeq3$ times lower than the JWST excess in Fig.~\ref{fig:SFR} due to more efficient UV emissivity). Importantly, more abundant massive stars boost the GRB formation efficiency. We may factor out the effect of the IMF on the efficiency as
\begin{align}
\eta_{\rm GRB}\propto\frac{\int_{m_{\rm GRB}}^{m_{\rm up}}\phi(m)dm}{\int_{m_{\rm low}}^{m_{\rm up}}m \phi(m)dm}\ ,
\end{align}
where $\phi(m)$ is the IMF defined for $m_{\rm low}<m<m_{\rm up}$, and $m_{\rm GRB}$ is the minimal mass to produce a GRB (we set $m_{\rm GRB}=25\,\Msun$ following \citealt{deSouza+2011}). For the Salpeter and top-heavy\footnote{\cite{Inayoshi+2022} considered a log normal distribution defined from $1\,\Msun$ to $500\,\Msun$ with a mean mass of $10\,\Msun$ dispersion of $\Msun$. In the estimation of $\eta_{\rm GRB}$, we excluded a mass window of pair instability supernovae, $140-260\,\Msun$ \citep[e.g.,][]{Heger&Woosley2002}.} IMFs in \cite{Inayoshi+2022}, this factor takes values of $\simeq1.4\times10^{-3}$ and $\simeq1.1\times10^{-2}$, respectively. Therefore, when the IMF completes the transition at $z\simeq11.3$, the GRB formation efficiency becomes $1.1/0.14\simeq7.9$ times higher than that at $z\simeq10$. This increase of $\eta_{\rm GRB}$ results in an overall increase of the GRB formation rate by $7.9/3\simeq2.6$ times higher than the case A beyond $z\gtrsim11.3$ (here the denominator, 3 comes from the reduction of the SFR from the value in case A). $\Psi_{\rm GRB}$ of this scenario are shown by dotted curves in Fig.~\ref{fig:Psi_GRB}. Note we simply interpolate $\Psi_{\rm GRB}$ in the transition period ($10\lesssim z \lesssim11.3$), although an actual shape depends on how the IMF changes over the period.

\item {\bf Case C: AGN.} Our last possibility is related not directly with star formation processes but other effects such as contribution from active galactic nuclei \citep[AGNs,][]{Harikane+2023b,Hegde+2024}. The corresponding GRB formation rates are actually already represented by the original GS22 (red solid) and $\eta_{\rm GRB}=\rm const$ models as the red and blue solid curves in Fig.~\ref{fig:Psi_GRB}. This is because the SFR nor GRB formation efficiency should not be increased to cause the JWST excess.
\end{itemize}

Note that while above three possibilities are listed up independently, they may be related with each other and interplay to shape the GRB formation rate. For instance, abundant massive stars born through a top-heavy IMF would cause an intense feedback diminishing the star formation efficiency \citep{Chon+2024,Menon+2024}. Nevertheless in this Letter we consider them separately as idealized cases and to see individual effects.

\section{Results}\label{sec:result}

\begin{figure}
\begin{center}
\includegraphics[width=85mm, angle=0, bb=0 0 277 320]{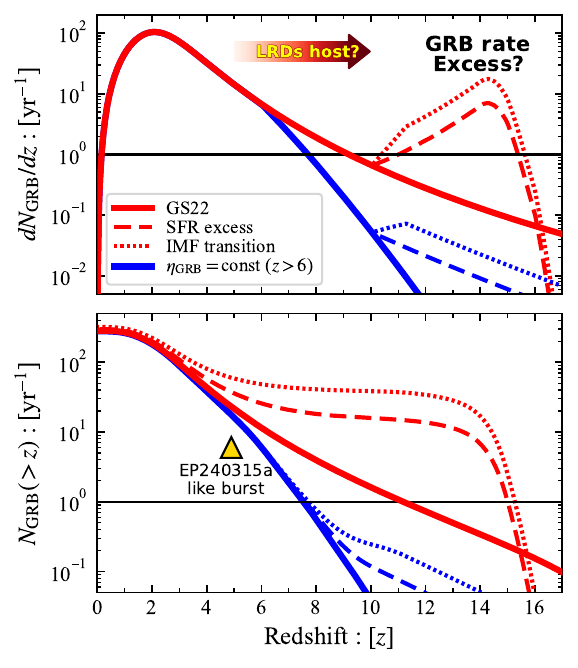}
\caption{({\bf Top}) Prediction for the redshift distribution of GRBs detected by EP. Color and types of curves correspond to the same meaning as in Fig.~\ref{fig:Psi_GRB}. The upper arrow represents a redshift range over which little red dots (LRDs) might be a host galaxy of GRBs, see discussion in Sec.~\ref{sec:summary}. ({\bf Bottom}) The cumulative number of GRBs detected by EP. The upward triangle means an expected lower-limit based on the detection of EP240315a.}
\label{fig:N}
\end{center}
\end{figure}

Figure.~\ref{fig:N} shows the redshift distribution and cumulative number of detected GRBs expected for the observation by EP. Within the models of GS22, the total number of GRB detection expected for one-year operation amounts to $\simeq280$ (see the bottom panel in Fig.~\ref{fig:N}), which could be larger than the actual number. As of September 1, 2024, $\simeq30$ X-ray transients have been reported on GCN circulars\footnote{\texttt{https://gcn.nasa.gov/circulars}}, corresponding to $\simeq50$ events per year, although currently not all detections are being reported and the instrument characteristics are still being investigated. In addition, our choice of the limiting flux, $\sim10^{-10}\,\rm erg\,s^{-1}\,cm^{-2}$, may not be applicable to compare our estimate with the current detection number since the observation by EP has just started and it underwent a commission phase. Another possibility is that a jet opening angle might be smaller than our fiducial value of $\theta_{\rm j}=0.1$. The detection of EP240315a at $z\simeq4.9$ only two months after the start of operation \citep{Liu+2024}, may also put a lower limit on the detection rate at $z\sim5$ (corresponding to 6 events per year, upward triangle).\footnote{If the first month after the launch of EP was not available for observation, the rate could be doubled.}

If the original GRB formation rate of GS22 can be extrapolated beyond $z\gtrsim6$ (red solid curve), a few GRBs at $z\geq10$ could be detected for one-year observation. The number of detection could be ten times larger if the JWST excess caused by the SFR excess or IMF transition; in these cases, one GRB could be detected per a few years even for the conservative model adopting $\eta_{\rm GRB}=const$ (blue solid curve). More specific detection numbers are shown in Table~\ref{table:result}.

\begin{table}
\begin{center}
\caption{Cumulative detection numbers of GRBs by EP in our each model.}
\label{table:result}
\begin{tabular}{lcccc}
\hline
&\multicolumn{3}{c}{Cumulative number}\\
Model&\multicolumn{3}{c}{$N_{\rm GRB}(>z)$ [$\rm yr^{-1}$]}\\
&$z=8$&$z=10$&$z=12$\\
\hline
GS22&4&2&0.7\\
\,\,\,\,\,+SFR excess&18&16&14\\
\,\,\,\,\,+IMF transition&41&39&34\\
$\eta_{\rm GRB}$=const ($z>6$)&0.6&0.04&0.002\\
\,\,\,\,\,+SFR excess&0.6&0.1&0.05\\
\,\,\,\,\,+IMF transition&0.8&0.2&0.1\\
\hline
\end{tabular}
\end{center}
\end{table}

More interestingly, the redshift distribution of GRBs can show an excess from an extrapolation from low redshifts or even increases for $z\gtrsim10$ depending on the origins of the JWST excess, which will be useful in discriminating the models. In case A (dashed curves in Fig.~\ref{fig:N}), the excess of the SFR also causes an excess of the GRB event rate in proportion to the SFR. In case B (dotted curves), the GRB rate increases more rapidly than case A during the period of the IMF transition ($10\lesssim z\lesssim11.3$). In case C (solid curves), the true SFR and GRB efficiency do not have an excess from the pre-JWST value, and hence the GRB rate extends smoothly as an extrapolation of the rate at $z\lesssim10$.

\begin{figure}
\begin{center}
\includegraphics[width=85mm, angle=0,bb=0 0 1022 661]{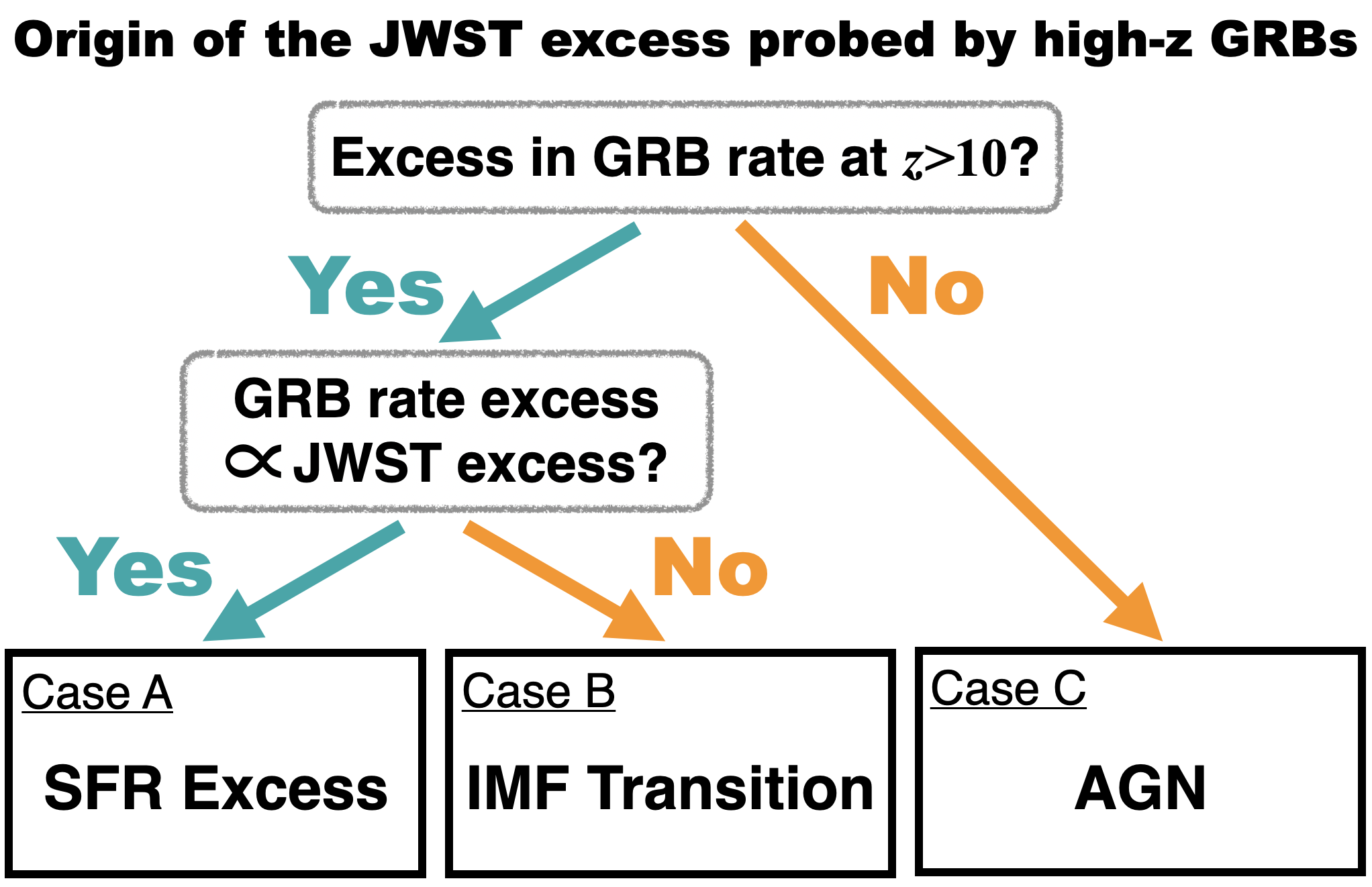}
\caption{Origin of the JWST excess implied by the behavior of the GRB event rate for $z\gtrsim10$. If the GRB rate shows an excess at $z\gtrsim10$ proportional to the JWST excess (the red or blue dashed curves in Fig.~\ref{fig:N}), an excess of the SFR is an origin of the JWST excess (Case A). If the GRB rate shows an excess larger than the JWST excess (the red or blue dotted curves), the IMF transition causes the JWST excess (Case B). If there is no excess in the GRB rate (the red or blue solid curves), AGN activity is responsible to the JWST excess (Case C).}
\label{fig:chart}
\end{center}
\end{figure}

Figure~\ref{fig:chart} summarizes the above discussion about a potential behavior of the high-$z$ GRB rate, and its implication on the origin of the JWST excess. If an excess in the GRB redshift distribution is identified at $z\gtrsim10$, it suggests that
the JWST excess is caused by a real elevation of the SFR from the pre-JWST value (case A) or a transition of IMF from Salpeter to top-heavy ones (case B). These scenarios could be discriminated by the amount of the increase in the GRB rate. No detection of an excess in the GRB rate supports that the JWST excess is caused by contribution from AGNs (case C).

\section{Discussion \& Summary}\label{sec:summary}

In this Letter, we explore the detectability of high-$z$ GRBs by WXT onboard EP, which started its operation in January 2024. Owing to the high sensitivity at the soft X-ray band, WXT is an ideal detector to access high-$z$ GRBs as demonstrated for EP240315a \citep{Liu+2024}. We find that EP could detect a few GRBs at $z\geq10$ for a one-year operation if the GRB formation rate calibrated by GRBs at $z\lesssim6$ can be extrapolated to $z\simeq10$. In particular, we focus on a synergy with JWST, which has recently reported an excess in the UV luminosity density and SFR density at $z\gtrsim10$ \citep[e.g.,][]{Harikane+2024b}. Since long GRBs are produced by collapse of massive stars, they probe the star formation activities in high-$z$ universe by directly tracing the star formation history. Interestingly, depending on potential origins of the JWST excess, the redshift distribution of GRBs shows different behaviors (Fig.~\ref{fig:N}). As summarized in Fig.~\ref{fig:chart}, if the JWST excess is caused by an elevation of the genuine SFR (case A), the redshift distribution has an excess at $z\gtrsim10$ in proportion to the JWST excess. If the transition of IMF from Salpeter to top-heavy one creates the JWST excess (case B), the distribution also shows an excess but the degree of the excess is different from case A. If other effects than star formation activities such as AGN contribution (case C), the distribution extends smoothly beyond $z\gtrsim10$.

It should be noted that EP alone, as an X-ray telescope, cannot determine redshift, and followups in optical and NIR wavelengths are critical to identify high-$z$ GRBs. Such followups may not be always possible for observation by EP, which may lower the number of detection than our estimates. Recently launched SVOM and future missions (see Table~\ref{table:detector}) with their own follow-up telescopes will play a role in the detection of high-$z$ GRBs \citep[e.g., see][for the prospect of SVOM]{LlamasLanza+2024}.

The nature of high-$z$ GRB's host galaxies is poorly understood, and their detection will be profitable \citep{Tanvir+2012b,McGuire+2016,Sears+2024}. Remarkably for most GRBs beyond $z>6$, only absorption by ISM in optical afterglow has been observed, and the emission from galaxies is hardly detected \citep{McGuire+2016}. We speculate that JWST-discovered puzzling galaxies, the so-called little red dots \citep[LRDs,][]{Harikane+2023b,Kocevski+2023,Labbe+2023,Matthee+2024} might be host galaxies of GRBs. LRDs are characterized by an extremely red rest-optical SED and compact size, and their origin is a mystery. Currently discussed possibilities include dust-obscured star-forming galaxies and AGNs. We propose that LRDs may represent a non-negligible fraction of high-$z$ GRB hosts, if they are star-forming galaxies. Some SED modeling of LRDs finds that the SFR is as high as $\sim10^{2-3}\,\rm \Msun\,yr^{-1}$ \citep{Xiao+2023,PerezGonzalez+2024}, which is 10-100 times higher than typical galaxies ($1-10\,\rm \Msun\,yr^{-1}$). Although the number density is $\sim10^{-2}$ times lower than normal Lyman break galaxies \citep{Greene+2024,Kocevski+2024,Kokorev+2024,Matthee+2024}, the potential high SFR suggests that they could at least partly contribute not only to the SFR density at a comparable level to normal galaxies \citep{Xiao+2023} but also to the GRB formation rate. GRBs hosted by LRDs would not show an afterglow in optical wavelengths shorter than $\lesssim0.73\left(\frac{1+z}{6}\right)\,\rm \mu m$ due to Lyman-$\alpha$ absorption, but they could be bright in NIR; hence JWST followup observations of ``dark'' GRBs \citep[e.g.,][]{Fynbo+2001b} with NIR counterparts may test this possibility.

Finally, detecting binary black hole (BBH) mergers at high redshifts might also be helpful to elucidate the origin of the JWST excess. BBH progenitors may experience GRBs \citep[e.g.,][]{Marchant+2016} while recent studies found that the fraction of GRBs evolving 
into a BBH detected by LIGO/Virgo is minor \citep[e.g.,][]{Arcier&Atteia2022,Wu&Fishbach2024}. However, since BBHs originate from massive stars, regardless of the connection with GRBs, their merger rate or mass distribution at high-$z$ may contain a hint to the JWST excess. In particular, unless the delay-time distribution is significantly shallow, the BBH merger rate would have an excess as in case A, and mass distribution would have a top-heavy shape in case B. Cosmic Explorer \citep{Reitze+2019} and Einstein Telescope \citep{Punturo+2010} will provide ideal opportunities to test these scenarios.

\begin{acknowledgements}
We thank Bing Zhang for stimulating discussion. We also thank the Yukawa Institute for Theoretical Physics at Kyoto University and participants to the YITP workshop YITP-W-24-22 on ``Exploring Extreme Transients: Emerging Frontiers and Challenges''. Discussions in the workshop were helpful to complete this work. This research is supported by the Hakubi project at Kyoto University, JSPS KAKENHI grant No. JP24K17088 (T.M.), JP24H00245, JP22K21349 (Y.H.), JP24H01810, JP24K00682, JP20H00174 (K.M.), MEXT/JSPS KAKENHI grant No. 23H01172, 23H05430, 23H04900, 22H00130, 20H01901, 20H01904, and 20H00158 (K.I.).
\end{acknowledgements}

\appendix
\section{Other GRB formation rate and LF}\label{Appendix:other model}
We briefly discuss the results for other GRB formation rate and LF. As other representative studies of the GRB population than GS22, we consider \cite{Lan+2021} and \cite{Palmerio&Daigne2021}. These works did not derive the intrinsic GRB formation rate $\Psi_{\rm GRB}$, but the rate for on-axis GRBs, which is expressed by $\eta_{\rm beam}\Psi_{\rm GRB}$ in our notation. 

\citealt{Lan+2021} analyzed 302 GRBs detected by Swift up to 2019 with a photon count rate greater than $\geq1\,\rm photon\,s^{-1}\,cm^{-2}$. They assumed that the (on-axis) GRB formation rate is proportional to the SFR as in Eq.~\eqref{eq:Psi_GRB}, and a broken power-law LF (Eq.~\ref{eq:LF_GS22}, defined for $10^{49}{\,\rm erg\,s^{-1}}<L<10^{55}{\,\rm erg\,s^{-1}}$). Several possibilities were considered such as combinations of both non-evolving GRB formation efficiency ($\eta_{\rm GRB}=const$) and LF ($k=0$), and either redshift-evolving efficiency or LF. We consider their result for the GRB formation rate with an evolving efficiency:
\begin{align}
\eta_{\rm beam}\Psi_{\rm GRB}=37.9(1+z)^{1.43}\left[\frac{0.0157+0.118z}{1+\left(\frac{z}{3.23}\right)^{4.66}}\right]{\,\rm Gpc^{-3}\,yr^{-1}}\ ,
\end{align}
where the bracketed factor comes from the SFR of \cite{Hopkins&Beacom2006,Li2008}, and the other factor shows the $z$-dependence of the GRB formation efficiency. The LF is assumed to be independent of redshift ($k=0$) and its parameters are obtained as $p_1=0.60$, $p_2=1.65$, and $L_{*}=10^{52.98}\,{\rm erg\,s^{-1}}$. 

\citealt{Palmerio&Daigne2021} modeled the same GRB sample as \cite{Ghirlanda&Salvaterra2022}. They parameterized the GRB formation rate as:
\begin{align}
\eta_{\rm beam}\Psi_{\rm GRB}(z)=\widetilde{\Psi}_{\rm GRB,0}\begin{cases}
e^{az}&: z\leq z_{\rm m}\ ,\\
e^{bz}e^{(a-b)z_{\rm m}}&: z>z_{\rm m}\ ,
\end{cases}
\end{align}
whose functional form is motivated by the cosmic SFR. They considered a Schechter LF \citep[][defined for $L>5\times10^{49}\,\rm erg\,s^{-1}$]{Schechter1976} with redshift evolving break:
\begin{align}
\frac{dn}{dL}\propto \left(\frac{L}{L_{\rm b}}\right)^{-p}\exp\left(-\frac{L}{L_{\rm b}}\right)\ ,
\end{align}
where the break luminosity is defined by Eq.~\eqref{eq:LF_GS22_Lb}. They did not determine the strength of $z$-evolution of the break luminosity $k$ by fit, but considered cases of $k=0$, 0.5, 1, and 2. All cases give a reasonable fit, and we adopt the case of $k=1$ as a moderate value. In this case, the parameters are obtained as $\widetilde{\Psi}_{\rm GRB,0}=0.72\,\rm Gpc^{-3}\,yr^{-1}$, $a=1.2$, $b=-0.27$, $z_{\rm m}=2.1$, $p=1.47$, and $L_*=10^{52.9}\,\rm erg\,s^{-1}$.

\begin{figure*}
\begin{center}
\includegraphics[width=170mm, angle=0, bb=0 0 500 336]{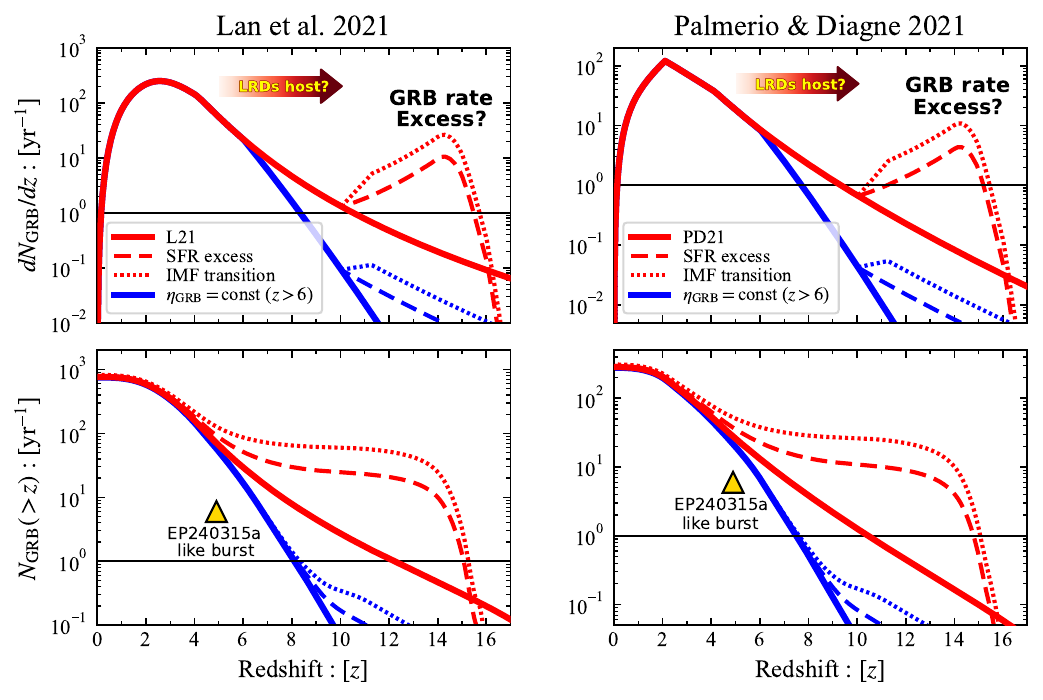}
\caption{The same as Figs.~\ref{fig:Psi_GRB} and \ref{fig:N} but for the GRB formation rate and LF of \citealt{Lan+2021} (L21, left) and \citealt{Palmerio&Daigne2021} (PD21, right), respectively.}
\label{fig:N_L21PD21}
\end{center}
\end{figure*}

Figure~\ref{fig:N_L21PD21} shows the redshift distribution and cumulative number of GRBs detected by EP for the cases of \cite{Lan+2021} and \cite{Palmerio&Daigne2021}. While the adopted GRB formation rate and LF are different from GS22, the results are qualitatively similar. The absolute values of detection rate for \cite{Lan+2021} is $\sim10$ times larger than GS22 and \cite{Palmerio&Daigne2021}, and it will be easily constrained by observations. The agreement between GS22 and \cite{Palmerio&Daigne2021} might support our choice of the opening angle of $\theta_{\rm j}=0.1$ if \cite{Lan+2021} is excluded in the future observations.

\bibliographystyle{aasjournal}
\bibliography{reference_matsumoto}

\end{document}